\documentclass[10pt,conference]{IEEEtran}

\usepackage[numbers]{natbib}
\usepackage{graphicx}
\usepackage{tabularx}
\usepackage{float}
\usepackage{hyperref}
\usepackage{longtable}
\usepackage{multirow}
\usepackage{array}
\usepackage[justification=centering]{caption}
\usepackage{tikz}
\usetikzlibrary{positioning}
\usepackage{soul}
\usepackage{hyperref}

\begin{document}

%% Paper title.
\title{The Dilemma of Privacy Protection for Developers in the Metaverse}

\author{
Argianto Rahartomo\thanks{argianto.rahartomo@tu-clausthal.de}\\
\scriptsize
Technische Universität Clausthal\\
\scriptsize
Germany
\and 
Leonel Merino\thanks{leonel.merino@uc.cl}\\
\scriptsize
School of Design, School of Engineering\\
\scriptsize 
Pontificia Universidad Cat\'{o}lica de Chile, Chile
\and 
Mohammad Ghafari\thanks{mohammad.ghafari@tu-clausthal.de}\\
\scriptsize
Technische Universität Clausthal\\ \scriptsize
Germany
\and
Yoshiki Ohshima\thanks{yoshiki.ohshima@acm.org}\\
\scriptsize
Shizuoka University\\
\scriptsize
Japan
}

\maketitle

\begin{abstract}
To investigate the level of support and awareness developers possess for dealing with sensitive data in the metaverse, we surveyed developers, consulted legal frameworks, and analyzed API documentation in the metaverse.
Our preliminary results suggest that privacy is a major concern, but developer awareness and existing support are limited.
Developers lack strategies to identify sensitive data that are exclusive to the metaverse.
The API documentation contains guidelines for collecting sensitive information, but it omits instructions for identifying and protecting it.
Legal frameworks include definitions that are subject to individual interpretation.
These findings highlight the urgent need to build a transparent and common ground for privacy definitions, identify sensitive data, and implement usable protection measures.
\end{abstract}


\section{Introduction}

The metaverse is recognized as an emerging and paradigm-shifting notion within the realm of digital technology.
It provides a virtual immersive environment in which humans can live, socialize, play, and work together or with virtual entities.
Today, we are closer to experiencing such a ``second life'' through virtual reality (VR), augmented reality (AR), and mixed reality (MR), all summarized in the concept of extended reality (XR).
Prominent platforms such as \emph{Second Life} and \emph{Roblox} have pioneered providing immersive virtual experiences and have been recognized as part of the metaverse ecosystem.
However, the metaverse has now evolved beyond these initial platforms, attracting large technology companies such as Microsoft, Apple, Facebook, and NVIDIA~\cite{wisnu_buana_metaverse_2023}.
According to predictions, annual global spending by companies and consumers related to the metaverse could reach \$5 trillion by 2030~\cite{dwivedi_wednesday_2023}.
One notable example is Snapchat, a popular social media platform recognized for its AR feature.
It allows users to apply face and background filters to their photographs and videos in real-time.
Snapchat has 414 million active users daily and reported USD \$4.6 million revenue only in 2023~\cite{snap2023}.

There is no ``one'' metaverse. The metaverse encompasses a collective virtual universe or a network of interconnected virtual environments, each with its own unique characteristics, rules, and user experiences~\cite{mystakidis_metaverse_2022}.
Generally, each metaverse obtains input from the physical world (users and environment) through interfaces such as cameras and sensors. 
Subsequently, it generates virtual entities such as avatars, digital objects, and metaverse services, all of which become perceptible through XR headsets and mobile applications to facilitate an immersive experience~\cite{Wu2023}.
%\rC{Do the authors consider a phone to be an XR headset? Most widely used AR, such as Snapchat, is mobile phone based.} 
%
In this paper, we consider metaverse applications beyond popular virtual worlds; we include augmented reality, worlds that mirror the physical world, and sensors technologies that report intimate states and life histories of objects and users.

The metaverse opens up many opportunities to live a digital life.
However, the significant amount and variety of personal data involved in metaverse experiences (e.g., biometrics) could also pose distinctive and substantial threats to users' privacy~\cite{chen_metaverse_2022}.
We wonder how metaverse developers are addressing this challenge and what support they have available.
%
% We draw the community's attention to a lack of privacy awareness and support in the metaverse.
%
In particular, we investigate the following research questions:

\vspace{0.3cm}
\noindent \emph{RQ.1 How aware are developers and users of privacy concerns in the metaverse?}

\noindent \emph{RQ.2 What assistance can developers find in technical documentation and legal frameworks?}
% \emph{RQ. What level of support and awareness do developers possess for dealing with sensitive data in metaverse?}\rA{It would be nice to "dropping down a level" to refine the overarching RQ into two, three or four subsidiary questions to frame what follows what give rise to a more convincing narrative}
\vspace{0.3cm}

We conducted a pilot study to investigate these research questions using a mixed-method approach.
We surveyed metaverse developers to learn about their understanding of sensitive data and their practices for privacy protection.
In addition, we complemented the survey's findings by interviewing metaverse end users to learn about their perceptions of privacy issues in this environment.
We consulted definitions of data privacy protection in legal frameworks, and looked at the documentation of popular immersive development toolkits.

None of the surveyed developers knew of a metaverse-specific data protection strategy.
For instance, they assumed that the captured spatial data are already anonymized by toolkits or that protection measures will be employed by the co-workers at the database level.
The results suggest that developers expect documentation of immersive development toolkits to include support for data protection.
We confirmed the gap existing between the broad scope of sensitive data definitions in legal frameworks and their applicability in concrete metaverse scenarios.
That is, developers with technical knowledge may not have a proper understanding of the legal frameworks.
We learned that metaverse users understand the meaning of sensitive data but are not fully aware of the potential privacy threats in the metaverse. 
Nevertheless, we also found that technical documentation supports only general recommendations, and actionable advice, such as code examples, is absent. 

In summary,
this work draws the community's attention to a lack of privacy awareness and support in the metaverse.
The dataset with interview and survey results is publicly available.\footnote{\url{https://doi.org/10.5281/zenodo.13138347}}

The remainder of this paper is organized as follows.
Section~\ref{sec:relatedwork} 
presents the motivation driving our study and an overview of related works.
In Section~\ref{sec:privacyawareness}, we discuss privacy awareness. 
Section~\ref{sec:privacysupport} examines technical documentation and legal frameworks.
We discuss threats to validity in Section~\ref{sec:threatstovalidity} and discuss the results in Section~\ref{sec:discussion}.

\section{Related Work}
\label{sec:relatedwork}
The tremendous impact of security and privacy risks in today's digital world seems to have ignited the adoption of better software development practices in the wild.
For example, a study of 182 open-source projects revealed that practitioners reported security issues in earlier development phases~\cite{Buhlmann22}.
However, security risks are still present even in critical systems, where security is a top priority.
To illustrate the matter, consider the Operational Technology (OT) that monitors and controls devices, processes, and infrastructure, and is used in industrial settings.
Despite a decade of efforts to improve security in this domain, the OT installation base still suffers from basic design issues such as hardcoded secrets~\cite{Wetzels23}. Alarmingly, these issues are even present in security-certified OT products.
Consequently, security and privacy are a more pressing need for emerging technologies.
Consider WebAssembly, which is a popular compilation target for running code on browsers and other platforms.
Several classic vulnerabilities such as stack smashing that are no longer exploitable in native binaries due to compiler protection measures may introduce risks when compiled to WebAssembly~\cite{Stivenart22}.
Moreover, developers often have limited awareness of the technologies available to secure user data privacy.
When they are aware of such technologies, they have to deal with the increased development cost of adopting them~\cite{boteju2023}.

The metaverse is an emerging technology that presents significant challenges in security and privacy.
Essentially, it involves numerous devices that collect extensive data in real time~\cite{abraham2022implications,de2019security,park2022, qamar_systematic_2023}.
Within this environment, users face the risk of sharing personal information through interactions with others.
Both end users and developers must grasp the importance of sensitive data.
A comprehensive examination of end-user perceptions and the assistance available to developers through legal frameworks and API documentation is necessary, especially to assess their effectiveness in safeguarding user privacy.
In the following, we analyze relevant previous studies that have investigated various aspects. 

Specifically, AR/VR technologies are susceptible to side-channel attacks, revealing user interactions (e.g., voice commands and keystrokes) along with information about the user's environment~\cite{zhang2023s}.
There are well-known risks and defenses for VR environments~\cite{munilla2024}.
In addition, interactions with virtual content expose users to certain risks. Notable instances include click-jacking attacks on Apple's ARKit, input-forgery incidents on Google's ARCore, and object-in-the-middle attacks on Meta's Oculus~\cite{cheng2024}.
Even movements of the user's head while typing on a virtual keyboard provide enough information to individualize a user, as demonstrated on the Meta Quest 2 system keyboard~\cite{carter2023}.
The memory artifacts of the HTC Vive headset can be used to recreate the immersive environment of the user, including its location, body positions, and the setup of the immersive room~\cite{CASEY2019S13}.
To address these challenges, significant initial steps have been taken to safeguard user data in the metaverse.
For example, applications could offer incognito modes to conceal sensitive information to prevent users from being profiled and losing anonymity~\cite{nair2023}.
Sensitive biometric information can be secured through approaches such as \emph{PrivXR}~\cite{warin2024}.
In addition, multiple privacy protection methods (e.g., federated learning, differential privacy, homomorphic encryption, and zero-knowledge proofs) could be adopted in software development; however, they are not used consistently.
By integrating these technologies, metaverse applications could effectively embed real-time privacy safeguards.

Legal frameworks serve as a source of guidelines for safeguarding sensitive data, yet developers encounter difficulties in comprehending these guidelines.
They need to translate legal language into technical implementations, which can be difficult as the two are not always aligned~\cite{marissa2023}. 
There are some solutions to support the implementation of GDPR requirements at different levels of software development~\cite{saltarella2021}.
However, verifying the compliance of developed applications with data privacy regulations is still a challenge~\cite{mcconkey2024}.
In effect, regulations like GDPR in metaverse environments may require expanding our view of human rights to address the datafication of both our physical and virtual realms.
This includes a wider interpretation of the right to freedom of thought to ensure mental autonomy, extending personality rights to avatars, and safeguarding rights to experiential authenticity and emotional and behavioral privacy.
In addition, neuro rights regarding physical and mental integrity and the safeguarding of brain data are essential.
% \rC{are they referring to consumer-based EEG systems? Can the authors be more specific?}
%
These concerns require a detailed discussion on the feasibility and necessity of such measures~\cite{hine2024}, asking whether developers understand these frameworks and can connect these guidelines to the requirements of the metaverse.

In this work, we conducted a pilot study to investigate the understanding of data privacy in the metaverse and the support available to developers.
We built on the results of valuable efforts that addressed privacy in metaverse~\cite{martin2022privacy,basyoni2024navigating,wang2023survey,huang2023security} but concentrate on linking the perceptions and the support provided to developers concerning legal and technical references, which we complement with end-user perception.
%
%\rB{Elaboration is quite preliminary yet. Even from a practitioner point of view, participants’ knowledge might be biased.}
%
This is an initial step towards inspiring our community to empower developers in identifying sensitive data and implementing resilient data protection mechanisms in the metaverse.

\section{Privacy Awareness}
\label{sec:privacyawareness}
%\rD{The extremely low number of questionnaire submissions makes any statistical analysis of the answers pointless.
%
%Likewise, the small number of answers precludes
%drawing any conclusions regarding the representativeness of the
%answers.
%
%Also, unfortunately also no statistical "proof" that the
%gut feeling is grounded}
%
To answer RQ.1, we investigated the extent to which developers and users are conscious of privacy issues within the metaverse.
Specifically, we surveyed developers actively engaged in metaverse technologies and interviewed end users of metaverse applications to understand their awareness of privacy. 
We received ethical approval for this study from our local university.\footnote{\url{https://doi.org/10.5281/zenodo.11200561}}
\subsection{Metaverse developers} 
We piloted a survey with three experienced developers.
We tested various questions and selected the ones that were more suitable for evaluating privacy awareness. 
We surveyed 14 developers, selected by convenience sampling.
These developers had extensive experience with immersive technologies and were based in prestigious organizations in Germany, New Zealand, USA, and Chile.
Each developer responded to an online questionnaire that included the questions listed in Table~\ref{tab:survey}.
They work on developing various types of metaverse applications, involving virtual and augmented reality.
These experts' insights helped reveal their understanding of sensitive data, its presence in the metaverse, and the protective measures that developers employ.
\paragraph{Demographics}
We asked participants to share their demographic information, specifically their level of education, age, gender, and occupation.
The average age was 34.7 \raisebox{.2ex}{$\scriptstyle\pm$} 11.6 years, highly educated (i.e., 86\% have an M.Sc. or Ph.D. degree). 
Of the 14 participants, 10 participants were from academia (that is, three professors, two postdoctoral researchers, and five master and doctoral students) who had hands-on experience with XR, and four were from industry (i.e., one data scientist, one software developer, and two video game engineers).
In general, participants had an average experience developing immersive applications of 7.4 \raisebox{.2ex}{$\scriptstyle\pm$} 8.1 years. 
\paragraph{What data is sensitive?}
Participants defined sensitive data using words such as \emph{private}, \emph{confidential}, \emph{control}, and \emph{protected}. 
Two reflected that sensitive data can potentially harm users' privacy, put people in danger, or be used to discriminate or violate fundamental rights. 
Three also described sensitive data as the ones that users do not want to share or would only share with specific people. 
Ultimately, we observe that they can describe the implications of misusing sensitive data, but they do not have strategies to identify them.
We asked participants about their awareness of the support provided by legal frameworks on data privacy management. 
Five participants identified GDPR as the most well-known legal framework. 
Other legal frameworks are the DSGVO (German Datenschutz-Grundverordnung) and the Chilean data protection law, each mentioned by a single participant. 
However, their knowledge was very shallow, and they were unaware of concrete definitions and requirements for protecting sensitive data. 
5
They rated their knowledge as moderately satisfactory, with a median score of 2.1 \raisebox{.2ex}{$\scriptstyle\pm$} 1.1.
Only one participant reflected that metaverse data are more detailed and abundant than in regular software applications. 
The rest of them did not recognize (and address) domain-specific needs in the metaverse.
Three participants mentioned that the strategies used to identify and protect sensitive data do not differ from the development process in other domains. 
Therefore, strategies are needed to protect these particular data.
\begin{table}
% [ht!]
\centering
% \small 
\caption{Developer-specific survey questions.}
\label{tab:survey}
\begin{tabular}{cp{7cm}} \hline
\textbf{\#} & \textbf{Question} \\ \hline
D1 & How long have you been developing immersive applications (e.g., \emph{virtual, and augmented reality})?\\
D2 & Please define "sensitive data" with your own words.\\
D3 & How important do you think is sensitive data protection? (Scale 1: \emph{Not too important} to 5: \emph{ Very important})\\
D4 & What would be the most sensitive information in your opinion?\\
D5 & How do you identify and protect sensitive data during software development?\\
D6 & How well (from 1 to 5) do you know about the legal frameworks that aim to protect our sensitive data? Name the frameworks and add your descriptions.\\
D7 & Do you think that sensitive data protection is a challenging concept in the metaverse? To what extent does this environment impose new challenges?\\
D8 & What features to protect sensitive data are available in extended reality platforms?
For example, \emph{ARKit}, \emph{ARCore}, \emph{MS HoloLens}, \emph{Windows Mixed Reality}, \emph{Magic Leap - Lumin}, \emph{Oculus}, and \emph{PlayStation VR}.\\\hline
\end{tabular}
\end{table}
\paragraph{Metaverse-specific sensitive data}
One participant mentioned that there are significant aspects, such as the sensors of the devices, through which information can be obtained without the user's awareness. 
Another identified multiple examples of personal data, for example, motion, face, eye tracking, and speech. 
In addition, they identified behavioral data, such as visited virtual locations and interactions with others (e.g., hanging out with specific communities), which can help reveal more information about the individual. 
Two also mentioned information about the surroundings (e.g., video and audio of home and family, the size of the play space, and the condition of the room) and personal preferences reflected in the avatars' designs (e.g., size, skin color, symbols, gender identity). 
One participant reflected that although some of the data might be kept locally or obfuscated, many are required to be transmitted to companies to personalize the virtual experience. 
A participant said that with this information, the systems can identify personal aspects, such as users' heights, speeds, or physical disabilities. 
One indicated that the metaverse also involves the creation of virtual identities that need to be protected to prevent acts such as identity theft or unauthorized access to sensitive information. 
Another participant foresaw that data theft could occur not only by nonauthorized access to computers but also directly to sensor devices. 
One identified sensitive information according to the guidelines. 
Another identified data protection needs by eliciting user requirements and collecting recommendations from ethical committees. 
Another participant adopted best practices found in scientific articles.
Although the participant did not provide details about the specific practices, the best practices referenced include data encryption methods, stringent access controls, and frequent security audits to safeguard data privacy and security in the metaverse.
We observed that they can recognize sensitive data in theory but lack strategies and tools to identify them in practice.
Participants demonstrated a foundational understanding of what constitutes sensitive data within metaverse, such as personal identifiers or location information.
However, when probed to explain the specific strategies or tools they deploy to identify and safeguard this data, their responses were generally vague or non-committal.
For instance, one participant acknowledged the importance of data encryption but admitted to being unfamiliar with the specific encryption protocols or software applications implemented in their projects.
They noted, ``\textit{Personally, I don't know what features those extended reality platforms have to protect sensitive data, since in most cases, that I have implemented, the data protection features come from other elements in the software system architecture, such as the database or the API}''.

% \rA{MG still does not understand the part ``the unique difficulty faced by metaverse developers lies in their capacity to deeply influencing users' intimate experiences by using the extensive personal data collected from them''.\\
% what does ``their'' refer to? how collected data impact user's intimate experience?
% }
% \ins{Although similar data-related challenges exist in other applications, the unique difficulty faced by metaverse developers lies in their capacity to deeply influencing users' intimate experiences by using the extensive personal data collected from them. This challenge is amplified by the heightened levels of immersion and presence promoted in metaverse environments, making users more susceptible to these effects, and making data protection a must.}

\paragraph{Protecting sensitive metaverse data}
%\rA{The nature of the data can be found in other applications, so the uniqueness of the challenges faced by metaverse developers doesn't come through at all}
%
All practitioners surveyed consider protecting sensitive data an important concern and have a basic understanding of sensitive data (with a median of 5 \raisebox{.2ex}{$\scriptstyle\pm$} 0.5 indicating its high importance).
Two protected sensitive data through encryption, authentication methods, and role-based access.
Four participants avoided collecting sensitive data and, when that was not possible, used anonymized data, but did not give details on how they ensured anonymity. 
Two relied on data protection features in the architecture of the software system, such as the database outside immersive development toolkits (e.g., ARKit, ARCore). 
One participant expected that sensitive data, such as the one scanned from the physical scene, would automatically anonymize using immersive development toolkits.
However, they were not sure if that happened. 
Another participant emphasized the importance of careful interaction design to enable users to handle sensitive information in public and shared spaces, such as metaverse applications.
He stated that in a real-time multi-user space like the metaverse, certain events and virtual objects need to be shared with all participants simultaneously, while others (that involve sensitive items) should be kept secure, for instance, by storing them locally.
He also added that sensitive data involved in transactions, such as payments, requires metaverse developers to be knowledgeable about web technology security practices (e.g., CORS), which are frequently used by payment platforms, posing a higher barrier to privacy/security protection in the metaverse.
All in all, we observe that the developers surveyed have not verified whether there is any data protection support in such toolkits.
Unfortunately, individual measures and the absence of standards could corrupt sensitive data~\cite{wang2022}, making it unusable when needed for auditing purposes~\cite{cloete2020}.
%
%\ins{Inconsistent data management can reduce the effectiveness of privacy audits. Future audits may require transaction records, and without data protection standards, ensuring proper data security becomes more challenging.}
%
%\rC{Why would "individual measures and the absence of standards" corrupt sensitive data for auditing purposes? What data needs to be available for a privacy audit?}. 

The limited number of questionnaire responses prevents us from performing statistical analysis. However, we believe these responses provide a valuable foundation for exploring the needs and support required to protect users' privacy in the metaverse.

\subsection{End-users perspective} 
We complemented our findings in the survey by interviewing end users within metaverse applications.
We aimed to know the user attitudes regarding sensitive data and privacy. 
We created avatars and visited various immersive applications such as Decentraland, Virbela, IMVU, Mozilla Hubs, and VRChat.
These avatars acted as a medium for social interactions, helping us to connect with other users in the virtual world. 
We approached the users of these applications and invited them to share their perceptions of sensitive data and privacy. 
We introduced ourselves and provided an overview of the research study's background before inquiring about their interest in participating in a conversation.
Upon their agreement, we then asked a set of demographic questions (i.e., level of education, age, gender, and occupation) to confirm that they are adults and can respond to our questions.
We designed semi-structured interviews to guide the conversation using the questions listed in Table~\ref{tab:interviews} and conducted them through both text and voice formats.

\begin{table}
\centering
\setlength\tabcolsep{4pt} % Adjust column spacing if needed
\caption{End-user specific interview questions.}
\label{tab:interviews}
\begin{tabular}{cp{7cm}} 
\hline
\textbf{\#} & \textbf{Question} \\ \hline
Q1&	What does sensitive data mean? Define in your own words and give me examples  \\
Q2&	How important is protecting your sensitive data for you? (Likert scale 1: not much to 5: very much)  \\
Q3&	Have you ever had any concerns regarding your sensitive data protection when using a metaverse application?  \\
Q4&	Can you tell us more about your concerns, examples of sensitive data, and the context?  \\
Q5&	How did you clear your concerns? Can you tell us about a solution to protect your data? \\
Q6&	Which of the following information do you consider sensitive data and wish not to disclose? (Likert scale 1: not much to 5: very much): 1) \textit{assets} (e.g., economic, email/mail content, photographs, physical, text message), 2) \textit{background} (e.g., credit information, criminal records, cultural, employee rec. inf., history of the Internet browser, parent names, place \& date of birth, product purchases records.), 3) \textit{identification} (e.g., ID number, name, online identifier, signature, social identity, social security number.), 4) \textit{location} (e.g., address, IP address, location data, phone number), and 5) \textit{medical} (e.g., genetic, health data, mental, physiological)  \\
Q7&	Is there any "\textit{other}" data that you consider sensitive?\\ \hline
\end{tabular}
\end{table}

%\begin{table}
% [ht!]
%\centering
% \small 
%\setlength\tabcolsep{2pt}
%\caption{End-user specific interview questions.}
%\label{tab:interviews}
%\begin{tabular}{lp{7cm}} \toprule
%\textbf{\#} & \textbf{Question} \\ \midrule
%Q1&	What does sensitive data mean? Define in your own words and give me examples  \\
%Q2&	How important is protecting your sensitive data for you? (Likert scale 1: not much to 5: very much)  \\
%Q3&	Have you ever had any concerns regarding your sensitive data protection when using a metaverse application?  \\
%Q4&	Can you tell us more about your concerns, examples of sensitive data, and the context?  \\
%Q5&	How did you clear your concerns? Can you tell us about a solution to protect your data? \\
%Q6&	Which of the following information do you consider sensitive data and wish not to disclose? (Likert scale 1: not much to 5: very much): 1) \textit{assets} (e.g., economic, email/mail content, photographs, physical, text message), 2) \textit{background} (e.g., credit information, criminal records, cultural, employee rec. inf., history of the Internet browser, parent names, place \& date of birth, product purchases records.), 3) \textit{identification} (e.g., ID number, name, online identifier, signature, social identity, social security number.), 4) \textit{location} (e.g., address, IP address, location data, phone number), and 5) \textit{medical} (e.g., genetic, health data, mental, physiological)  \\
%Q7&	Is there any "\textit{other}" data that you consider sensitive?\\ \bottomrule 
%\end{tabular} \end{table}

\paragraph{Results}
In total, 11 people accepted our invitation to share their opinions on the meaning of sensitive data and privacy concerns in the metaverse.
These people had various occupations (e.g., pharmacists, engineers, and students). 
They corresponded to young adults between 21 and 32 years of age ($\mu=$ 26) and had used the metaverse for one to four years.
Four of the interviewees were women.
The interviewees unanimously defined sensitive data as \emph{that can reveal their real identity} and mentioned traditional examples of sensitive data such as identification number, name, phone, postal address, email address, employment information, bank account, fingerprint and voice.
Two interviewees mentioned that an avatar that assimilates their characteristics in real life (e.g., outfit, physical features, name, actions) might reveal their identities.
Unexpectedly, the interviewees did not mention any other metaverse-specific risk such as environmental data.
However, the interviews revealed that the participants are very concerned about their privacy and prefer to have a ``second identity'' in the metaverse and stay anonymous.
Users should know the limits to their privacy to protect their real identities in the metaverse.
Otherwise, due to the abundant sensitive information that is collected in this domain and the blurry distinction between real and virtual, engaging in tactics such as assuming a second identity is very vulnerable.
\section{Privacy Support}
\label{sec:privacysupport}
To examine RQ.2, we analyzed the privacy support available to developers and end users. Specifically, we reviewed the technical documentation of popular metaverse frameworks and consulted the main legal frameworks that focus on data privacy management.
In the latter two cases, two authors conducted this analysis separately, and their findings were consistent.

\subsection{Technical Documentation}
Several platforms exist that support developers to create immersive applications, e.g., Google ARCore, Apple ARKit, Microsoft HoloLens/Windows Mixed Reality, Magic Leap, OpenXR, OpenVR, Oculus, and Viveport SDK.
We inspected each platform's documentation to analyze the support developers can find in toolkits for identifying and protecting sensitive data. 
We searched pages that contain terms such as ``sensitive'', ``privacy'', or ``security'', as they are the most commonly mentioned terms in developer discussions related to security and privacy~\cite{tahaei2020understanding}.
We checked whether every page containing these keywords included any information for security and privacy protection.
Regarding toolkits that did not include pages specifically mentioning these keywords, we examined the documentation's structure and identified sections that discussed data management support.
We selected two popular toolkits, namely ARCore and ARKit, because we had similar observations across different platforms.
\paragraph{ARCore} 
Google Play Services for AR (ARCore) collects the following user data: device IDs, performance and diagnostic data, and API usage and app activity.
Such data is not shared with third parties and are stored by Google. 
Additional data is also collected when apps use particular APIs (i.e., ARCore Geospatial API and Cloud Anchors).
In shared AR experiences, the platform requires an understanding of users' physical environments to allow multiple users to view and interact with virtual objects in a shared physical space. 
To this end, it uploads images from the user device's camera to Google servers.
The platform protects users' privacy by deleting images after processing and securely storing visual data (which are not accessible to developers).
The documentation indicates that applications must disclose the use of ARCore, Cloud Anchors, and ARCore Geospatial API, request camera permissions, and notify users that Google is processing visual data from the camera.
Developers are also allowed to record sensitive data, such as video, sensors, and custom data.
For instance, developers can instrument apps to record the user's living room so users can add virtual furniture on the go.
We could find ample technical examples that guide developers in collecting such data, but we have not come across \emph{any instructions} on how to adequately identify and protect sensitive information.
\paragraph{ARKit} Developers are allowed to detect and track faces, capture body motion, and track people in their physical environment and their proximity to anchor objects. 
The documentation of Apple's AR platform only informs developers to ask users' permission to access the camera and to include a privacy policy describing how data will be handled.
We did not find any of our ``security'', ``sensitive'', or ``privacy'' keywords in the documentation. 
Only a data management section with guidelines for managing facial and body data exists.
Rather, it explains technical capabilities without warning developers that sensitive data must be protected.
In general, we discovered several APIs that operate on sensitive user data, but the documentation was mostly limited to warnings such as ``session recordings may contain sensitive information'' without any examples or references to security measures.
Unfortunately, it is not possible to determine beforehand whether a recording includes sensitive information, and there is no support from the framework to detect and protect if sensitive data are present. 
Indeed, identifying sensitive data and uncovering which APIs may handle such data in an extremely dynamic environment like the metaverse is challenging.
\subsection{Legal Frameworks}
Legal frameworks provide comprehensive definitions of sensitive and personal data. 
By adhering to these regulations, individuals can exert greater control over their personal information, fostering digital trust and privacy.
To understand the definitions and implications of personal and sensitive data (\textit{sometimes referred to as a special category of personal data}), we consulted the legal frameworks of multiple regions of the world listed.
Namely, Europe's General Data Protection Regulation (GDPR), California Consumer Privacy Act (CCPA), Australia's Privacy Act, China's Personal Information Protection Law (PIPL), South Africa's Protection of Personal Information Act (POPIA), Brazil's General Data Protection Law (LGPD), Japan's Act on the Protection of Personal Information (APPI), Turkey's Personal Data Protection Law (KVKK), India's Personal Data Protection Bill (PDPB), Indonesia's Personal Data Protection Act (UU PDP), and Chile's Protection of Private Life (DPL).
We started mainly with the article that contains the definition of sensitive data (for example, Article 9 in GDPR), and read any further references that were given (e.g., recital 46 in GDPR). 

Legal frameworks do not seem to be a good source of information for developers.
They include definitions that are subject to individual interpretations and external influences.
In addition, the lack of instruments and diverse interpretations makes measuring the compliance level a challenging task~\cite{ayala2018}. 
Hence, as one senior Metaverse developer stated, the industry often looks up minimal documentation online that explains \emph{"how to comply with the rules"} and then attempts to meet the requirements with minimal effort. This allows them to focus more on the delivery of the product.
\paragraph{Diverse definitions}
Although we did not notice conflicting definitions, we observed that the characteristics of sensitive data in various frameworks are diverse.
They share common ground but describe it with different levels of depth. 
In general, GDPR, PIPL, PA, and CCPA have more concrete definitions and examples than other frameworks.
People may have varying opinions about what constitutes sensitive data and interpret definitions differently. 
There could also be external influences (e.g., cultural, religious, economic, and social).
For example, in countries where oppressive regimes or political instability prevail, people may opt to hide their political beliefs and affiliations to protect themselves from persecution or discrimination.

\paragraph{Domain-specific sensitive data}
The sensitive nature of the data can be revealed in a specific context.
For example, none of the legal frameworks we inspected included \emph{"personal distance"} as a privacy concern, while \emph{"safety distance"} has become an important privacy need in the metaverse since women reported sexual abuse in this environment.\footnote{\url{https://nypost.com/2022/05/27/women-are-being-sexually-assaulted-in-the-metaverse}}
This highlights a gap in the way existing regulations address context-specific privacy needs.
Moreover, the low barrier for collecting massive behavioral data could enable machine learning techniques to synthesize biometric measures and reveal people's identities.

\paragraph{Legal-friendly Technical Framework Development}
In response to these evolving challenges, it is essential to create a dynamic technical framework that not only complements but also develops alongside legal standards to offer strong context-specific technical solutions~\cite{diepenbrock2023}. 
This framework must be consistently adjusted to integrate technological advancements, ensuring that privacy protection progresses with technological innovations.

We observed that legal frameworks represent the entry point for a requirements engineering process, which should lead to concrete security and privacy requirements for a system, possibly, based on a threat and risk analysis. However, such a process seems missing.

\section{Threats to Validity}
\label{sec:threatstovalidity}
%
%\rC{it is not enough to acknowledge the problem in the threats section. The authors should consider this study as a pilot and should be presented as such}

%
In this section, we explain several threats to the validity of this study.

\emph{Selection and Self-Selection Bias} Participants in our user interviews and developer surveys may not fully represent the diversity of metaverse users and developers.
Focusing on active users of specific platforms, potentially excluding those with different usage patterns, could lead to selection bias.
Moreover, self-selection bias could affect participants' willingness to engage based on their interest in sensitive data and privacy concerns.
We mitigated this threat by involving various metaverse communities.

\emph{Expertise Bias} Our metaverse developer survey mainly included experts with substantial experience and education.
This could lead to an overrepresentation of advanced viewpoints, neglecting perspectives from less experienced developers or those from different sectors.
To mitigate this threat, we include questions to discuss aspects that could also involve the viewpoints of less experienced developers.

\emph{Social Desirability Bias} Participants in user interviews and surveys could provide responses that align with social desirability, affecting the precision of their genuine attitudes and practices.
This bias might lead to an overestimation of the importance of privacy-related issues.
We mitigated this threat by explicitly stating the anonymity of the participants' responses.

\emph{Toolkit Documentation Scope} Analyzing toolkit documentation using specific keywords might overlook comprehensive guidance on sensitive data protection that does not explicitly use those terms.
This approach could limit our understanding of the support available to developers.
We mitigated this threat by cross-examining the results and reading the documentation.

\emph{Measurement Instrument Reliability} The survey questionnaire used for the developer survey might not cover the full range of sensitive data protection practices.
The design and options provided could influence how developers express their opinions, affecting the reliability of the data.
To mitigate this threat, we piloted the survey with a subset of developers and improved it with their feedback.
We also included open-ended questions to elicit a wide range of responses.
\section{Discussion}
\label{sec:discussion}
Privacy issues do not interrupt users and are not immediately visible to them; therefore, users may not be aware of potential privacy violations.
Quantifying privacy becomes even more challenging due to the vast amount of sensitive data shared in various contexts, such as education, business, and health, where the requirements and implications for data privacy can differ significantly.
These different contexts can shift security awareness and require context-specific protective measures.
We showed that developers do not possess specific strategies to identify and protect sensitive information.
There are no guidelines in the API technical documentation to ensure adequate data protection measures.
Protection measures in the development process are ad hoc if present at all. 
Developers usually focus on product delivery and adhering to legal standards with minimal effort.
Lack of standards can lead to information leakage or render sensitive data unusable where needed, such as for auditing.
We recommend that measures such as those implemented based on user-permission models on mobile apps would be helpful in the metaverse domain.

Legal frameworks also do not seem to be a good source of information for developers.
Existing definitions are high-level and subject to individual interpretation and external influences.
%
% Within a large company, developers may benefit from the availability of legal staff who can be consulted for privacy protection matters, but this is not the case for small companies and in-house developers. 
%
In large companies like Google, legal staff and privacy engineers often ensure compliance with regulations. However, their numbers are very small compared to the number of developers. In addition, small companies and internal developers typically lack access to such resources.
% Many companies off-load the responsibility of keeping features in-line with legal regulations with dedicated engineers who are not directly embedded into the project. Specifically, Google has approximately 500 "privacy engineers" whose job is to ensure compliance with regulations. 500 people is a very small number in comparison to ~20,000 software developers at the company.
%
We observe that the metaverse's evolving and dynamic nature makes it very difficult to determine which data are sensitive.
What was once not sensitive may reveal sensitive information in the future.
Similarly, while individual pieces of data may not appear sensitive, their combination has the potential to compromise privacy. 
The end-user perspective is also important.
Given the need for a multidimensional approach, privacy protection cannot solely depend on a single facet.
It is essential to consider integrating user-centric approaches with legal compliance.
It might be possible to explore the concept of using the incognito mode in the metaverse environment, similar to the incognito mode in web browsers, to protect privacy.
However, the metaverse differs in complexity compared to a web browser.

We believe that to maintain the metaverse as a realm of opportunity rather than susceptibility, it is necessary to develop
(1) techniques for identifying potentially sensitive data, 
(2) legal standards to prevent misinterpretations or assistive tools to support legal terms interpretation and legal compliance measurement,
and (3) mechanisms to protect sensitive data while allowing access when necessary, such as for auditing purposes.
Evaluating the necessity of all the data gathered by the hardware in the metaverse is crucial.
Large data collections, if handled improperly, can increase security vulnerabilities.
A better approach to data gathering and transfer processes should be considered.
These approaches must efficiently address potential security issues during data exchanges between entities.

We observe the need for increasing efforts on creating a transparent and common ground for privacy definitions in official documentation and technical resources. 
Developers should be able to determine if an API involves sensitive data, and documentation must explicitly specify when sensitive data are automatically protected or if that is the responsibility of developers.
We observe the need to implement privacy protections at the development level, potentially based on various user permissions, similar to practices in the mobile app domain.
The implemented measures should protect sensitive data and make them available in a meaningful format when needed for tasks like auditing.

\section{Conclusion}
\label{sec:conclusion}
The tremendous amount of sensitive data shared in the metaverse can significantly compromise people's security and privacy.
We conducted a pilot study to explore developers' and end-users' awareness of this issue and to assess whether adequate support exists for data protection.
In particular, we surveyed 14 developers actively engaged in metaverse technologies, looked at technical documentation and legal frameworks, and interviewed 11 end users of metaverse applications.
We observed that metaverse users understand the meaning of sensitive data but are unaware of potential privacy threats.
We learned that developers do not have specific strategies to identify and protect sensitive information.
Legal frameworks include definitions that are subject to individual (mis)interpretations. 

To reach a reliable conclusion, future studies with a larger sample size and thematic analysis are warranted.

% We observed the need for increasing collaborative efforts to develop a common ground for defining and protecting privacy in the metaverse.
%

\section*{Acknowledgment}
Leonel Merino acknowledges funding by ANID FONDECYT Iniciación Folio 11230349.

\bibliographystyle{ieeetr}
\bibliography{main}

\end{document}